\documentclass[a4paper,12pt]{article}
\usepackage{amssymb,amsmath}
\usepackage{a4wide}
\usepackage{epsfig}
\usepackage{graphics}
\usepackage{graphicx}
\usepackage{subfigure}
\usepackage{amsfonts}
\title{Deformed Skyrme Crystals}
\author{J.\ Silva Lobo\footnote{email address: j.i.silva-lobo@durham.ac.uk}
  \bigskip
  \\Department of Mathematical Sciences,
  \\Durham University,
  \\Durham DH1 3LE}
\date{}

\begin{document}
\maketitle
\begin{abstract}
The Skyrme crystal, a solution of the Skyrme model, is the lowest
energy-per-charge configuration of skyrmions seen so far. Our numerical
investigations show that, as the period in various space directions is changed,
one obtains various other configurations, such as a double square wall, and
parallel vortex-like solutions. We also show that there is a sudden "phase
transition" between a Skyrme crystal and the charge 4 skyrmion with cubic
symmetry as the period is gradually increased in all three space directions.  
\end{abstract}
\vskip 1truein

%\noindent {\bf Keywords:} Solitons Monopoles and Instantons; Sigma Models;
%   Global Symmetries. 

\newpage

%%%%%%%%%%%%%%%%%%%%%%%%%%%%%%%%%%%%%%%%%%%%%%%%%%%%
\section{Introduction}

The Skyrme model was originally proposed by Tony Skyrme in 1961
\cite{skyrme:1961}. It is a theory of nuclear matter in which the fundamental
building blocks are pion fields. It was later shown \cite{witten:1983,
witten:1979} that it can be regarded as a low-energy approximation to QCD, which
becomes more accurate in the large $N_{c}$-limit, where $N_{c}$ is the number of
quark colours. 
The (topologically non-trivial) solutions of this theory are called skyrmions.
There is a special way of arranging them in order to obtain the smallest known
value of the energy per baryon number seen so far \cite{mands}, known as the
Skyrme crystal. This is an infinite, triply-periodic, arrangement of
half-skyrmions
\cite{Klebanov:1985,Goldhaber:1987,Jackson:1988,Castillejo:1989,Kugler:1988,
Kugler:1989}. In order to study Skyrme crystals, we impose periodic boundary
conditions, with period $L$, on the Skyrme field in all three directions. The
skyrmions are therefore defined on a 3-torus $T^{3}$.

Skyrme crystals were originally proposed by Klebanov \cite{Klebanov:1985} as a
model for dense nuclear matter, such as that found in neutron stars. At the
time, the behaviour of two well-separated skyrmions was already known
\cite{skyrme:1961} -- an important feature being that these are
maximally-attracted when one is rotated with respect to the other by
$180^{\circ}$ about a line perpendicular to the line connecting them. Klebanov
believed that an interesting extension of this idea would be to have an array of
skyrmions, where any skyrmion would be attracted by its nearest neighbours. A
graceful way of achieving this, Klebanov showed, would be to arrange the
skyrmions in a simple-cubic lattice with appropriate rotations applied to all
the nearest neighbours of a chosen skyrmion. Moreover, if the skyrme fields are
to have the correct periodicity in all three directions, they must have symmetry
elements that combine both spatial transformations as well as isospin
transformations acting on the individual fields (more on this later in the
introduction).

Klebanov also showed that there is a minimum in the energy per baryon number of
the lattice for a certain value of the period $L$. In other words, there is a
preferred size of the fundamental cell. Later, Goldhaber and Manton showed
\cite{Goldhaber:1987} that there is a phase transition from a low-density
simple-cubic lattice of skyrmions to a high-density body-centred lattice of
half-skyrmions. However, it has since been shown by Kugler and Shtrikman
\cite{Kugler:1988,Kugler:1989} and by Castillejo et al. \cite{Castillejo:1989}
that the lowest energy per baryon configuration is that of a (high-density)
half-skyrmion phase corresponding to an initial (low-density) face-centred cubic
(fcc) array of skyrmions. 

An fcc array is one in which skyrmions with standard orientation are placed on
the vertices of a cube and more skyrmions are placed on the face centres, but
this time rotated by $180^{\circ}$ about an axis perpendicular to the face. Such
a configuration produces 12 nearest neighbours (to a particular skyrmion), which
are all in the attractive channel. If the origin is fixed at the centre of one
of the unrotated skyrmions and the skyrme fields are given by $\Phi_{\beta}{\bf
(x)}=(\Phi_{1}{\bf (x)}, \Phi_{2}{\bf (x)}, \Phi_{3}{\bf (x)}, \Phi_{4}{\bf
(x)})$, then the fcc configuration would have spatial and isospin symmetries, as
was alluded to above. The generators for these symmetries are listed in Table 1
\cite{mands}:

\begin{table}[ht]
\caption{Symmetry generators of an fcc array of skyrmions}
\centering
\scalebox{0.85}{%
\begin{tabular}{c c c}
\hline\hline
Transformation Name & Spatial Transformation & Isospin Transformation \\ [0.5ex]
\hline
$\mathcal{R}_{1}$ & $(x_{1},x_{2},x_{3}) \to (-x_{1},x_{2},x_{3})$  &
$(\Phi_{1},\Phi_{2},\Phi_{3},\Phi_{4}) \to
(-\Phi_{1},\Phi_{2},\Phi_{3},\Phi_{4})$ \\

$\mathcal{R}^{3}_{1,-}$ & $(x_{1},x_{2},x_{3}) \to (x_{2},x_{3},x_{1})$  &
$(\Phi_{1},\Phi_{2},\Phi_{3},\Phi_{4}) \to
(\Phi_{2},\Phi_{3},\Phi_{1},\Phi_{4})$ \\

$\mathcal{R}^{4}_{1,-}$ & $(x_{1},x_{2},x_{3}) \to (x_{1},x_{3},-x_{2})$  &
$(\Phi_{1},\Phi_{2},\Phi_{3},\Phi_{4}) \to
(\Phi_{1},\Phi_{3},-\Phi_{2},\Phi_{4})$ \\

$\mathcal{T}_{1+,2+}$ & $(x_{1},x_{2},x_{3}) \to (x_{1}+L/2,x_{2}+L/2,x_{3})$  &
$(\Phi_{1},\Phi_{2},\Phi_{3},\Phi_{4}) \to
(-\Phi_{1},-\Phi_{2},\Phi_{3},\Phi_{4})$ \\ [1ex]
\hline
\end{tabular}}
\label{table:gensymmetries}
\end{table}

The transformations listed here are $\mathcal{R}_{1}$: a reflection in the
$x_{1}-$axis, $\mathcal{R}^{3}_{1,-}$: a "negative" three-fold rotation about
the diagonal that goes from the origin to the opposite corner of the cube
(defined as "$1$"), $\mathcal{R}^{4}_{1,-}$: a "negative" four-fold rotation
about the $x_{1}-$axis, and $\mathcal{T}_{1+,2+}$: a positive $L/2-$translation
in both the $x_{1}-$axis and the $x_{2}-$axis. Note that these are a subset of
the possible transformations that can be carried out on an fcc lattice. For
example, one can also have a "positive" three-fold rotation along the same
diagonal, $\mathcal{R}^{3}_{1,+}$, given by the transformation:
$(x_{1},x_{2},x_{3}) \to (x_{3},x_{1},x_{2}), \qquad
(\Phi_{1},\Phi_{2},\Phi_{3},\Phi_{4}) \to
(\Phi_{3},\Phi_{1},\Phi_{2},\Phi_{4})$. Another possible transformation could
also be a "positive" four-fold rotation about the $x_{3}-$axis,
$\mathcal{R}^{4}_{3,+}$: $(x_{1},x_{2},x_{3}) \to (-x_{2},x_{1},x_{3}), \qquad
(\Phi_{1},\Phi_{2},\Phi_{3},\Phi_{4}) \to
(-\Phi_{2},\Phi_{1},\Phi_{3},\Phi_{4})$.

There is an additional symmetry unique to the high-density phase of
half-skyrmions. Its generator is given by:

\begin{table}[ht]
\caption{Additional symmetry generator of the high-density half-skyrmion phase}
\centering
\scalebox{0.85}{%
\begin{tabular}{c c c}
\hline\hline
Transformation Name & Spatial Transformation & Isospin Transformation \\ [0.5ex]
\hline
$\mathcal{T}_{1+}$ & $(x_{1},x_{2},x_{3}) \to (x_{1}+L/2,x_{2},x_{3})$  &
$(\Phi_{1},\Phi_{2},\Phi_{3},\Phi_{4}) \to
(-\Phi_{1},\Phi_{2},\Phi_{3},-\Phi_{4})$\\ [1ex]
\hline
\end{tabular}}
\label{table:gensymmetries}
\end{table}

Note that this transformation involves a chiral $\mathop{\rm SO}(4)$, rather than just isospin $\mathop{\rm SO}(3)$ rotations displayed in Table 1 above and it can replace the $\mathcal{T}_{1+,2+}$ transformation since that can be achieved through successive applications of $\mathcal{T}_{1+}$ and $\mathcal{T}_{2+}$.

In this letter, we will see that the field behaves in an interesting way as one
changes the period of the crystal in different ways, for different directions.
For example, if we start with the Skyrme crystal and then increase the period
along all three space dimensions in the same way, one gets the familiar picture
of the cubically-symmetric charge $Q=4$ skyrmion. We also show that the energy
density of the $Q=4$ skyrmion displays certain similarities previously seen in
the context of Skyrme chains, provided we have large $L_{x,y}$ (where
$L_{x}=L_{y}$ are the periods in the $x_{1}-$ and $x_{2}-$directions) and small
$L_{z}$ values. Skyrme chains are solutions of the Skyrme model, which are
periodic in one space dimension. It has been shown \cite{Harland:2008eu} that
soliton chains generally have constituents in the form of vortex-antivortex
pairs. These emerge when the period is small compared to the natural soliton
size - a feature that we verify when the period in the z-direction is small
compared to the other two space dimensions. When the period increases, the
constituents tend to clump together, a feature that is also verified here. 

A double Skyrme sheet \cite{jslrsw}, is also seen to emerge at small $L_{x,y}$
and large (and effectively infinite) $L_{z}$ values. It takes the form of a
square lattice, an object analogous to the hexagonal ``Skyrme domain wall''
solution \cite{Battye:1997wf}. However, our system is periodic in all three
directions, which means that the vacuum value on both sides of the Skyrme sheets
($\pm \infty$ in the z-direction) is unique.

Finally, we describe what happens as one increases the period simultaneously in
all three directions, starting with the Skyrme crystal, and show that there is a
rapid transition between the Skyrme crystal and the $Q=4$ skyrmion with cubic
symmetry. We show evidence which suggests that this is a second-order phase transition with
an order parameter given by the period $L_{x,y,z}$ of the configuration (where $L_{x}=L_{y}=L_{z}$).

%%%%%%%%%%%%%%%%%%%%%%%%%%%%%%%%%%%%%%%%%%%%%

\section{The Skyrme Crystal} \label{sec:llxyslz}
\subsection{Background} 
The static energy density of the Skyrme model is given by:

\begin{equation} \label{eq:skyrmelag}
  \mathcal{E}=-\frac{1}{2}\mathrm{Tr}(L_i L_i)-
     \frac{1}{16}\mathrm{Tr}([L_i,L_j][L_i,L_j])\,, 
\end{equation}
where $L_i=U^{-1}\partial U/\partial x^i$, $x^i=(x,y,z)$ are the spatial
coordinates, and the field $U(x^{i})$ is an SU(2)-valued scalar field. The pions
are grouped into this scalar field as follows:
\begin{equation}\label{eq:uaspions}
U = \Phi_{4} + i\Phi_{i}\sigma_{i}\,,
\end{equation}
where $\Phi_{i}$ is a triplet of pion fields, $\sigma_{i}$ is the triplet of
Pauli matrices, and $\Phi_{4}$ is an additional scalar field determined through
the constraint: $U^{\dagger}U=\Phi_{\beta}\Phi_{\beta}=1$.

The energy $E$ is defined as
\begin{equation}\label{eq:sken}
  E=\frac{1}{12\pi^2}\int \mathcal{E}\,dx\,dy\,dz\,.
\end{equation}

The scalar field $U$ is a map from $\mathbb{R}^{3}$, compactified at infinity,
to $S^{3}$, the group manifold of SU(2). Skyrme identified the degree of this
map, a topological invariant, with the baryon number, which is given by
\begin{equation}\label{skcharge}
  Q = \int \mathcal{Q}\, dx\, dy\, dz\,,
\end{equation}
where
\begin{equation} \label{skchargeden}
  \mathcal{Q}= \frac{1}{24\pi^2}\varepsilon_{ijk}\mathrm{Tr}(L_i L_j L_k)
\end{equation}
is the topological charge density. The energy (\ref{eq:sken}) satisfies the
Faddeev-Bogomolny lower bound \cite{faddeev} $E \ge Q$.

An analytic approximation for the fields of the Skyrme crystal was proposed in
\cite{Castillejo:1989}. It takes into account the $\mathop{\rm SO}(4)$ chiral symmetry as well as the $\mathop{\rm SO}(3)$ isospin symmetries. The fields are expressed as follows:
{\setlength\arraycolsep{2pt}
\begin{eqnarray}
\Phi_{4} &=& c_{1}c_{2}c_{3}, \label{eq:appsolcrystal1}\\
\Phi_{1} &=&
-s_{1}\left(1-\frac{s_{2}^{2}}{2}-\frac{s_{3}^{2}}{2}+\frac{s_{2}^{2}s_{3}^{2}}{
3}\right)^{\frac{1}{2}}
\textrm{  and cyclic permutations,} \label{eq:appsolcrystal2}
\end{eqnarray}}
where $s_{i}=\sin \left(2\pi x^{i}/L\right)$ and $c_{i}=\cos
\left(2\pi x^{i}/L\right)$. It is a good approximation to the actual
minimal-energy solution.

\subsection{Changing the periods $L_{x}=L_{y}$ and $L_{z}$}
In what follows, we take periodic boundary conditions in all three space
directions - the periods will be specified in the relevant sections. The lattice
spacings in the $x,y,\textrm{ and }z$ directions are given by
$h_{x},h_{y},h_{z}$ and the number of lattice points are given by
$n_{x},n_{y},n_{z}$, yielding side-lengths $L_{x,y,z}=h_{x,y,z}*n_{x,y,z}$. We
use a first-order finite-difference scheme and implement a full 3-dimensional
numerical minimization of the energy using the conjugate gradient method
(\emph{cf.} \cite{numrecipes}). 

Note that there is a numerical error associated with the finite lattice spacing.
The way we approximate the errors in the energy, for a given configuration, is
by comparing its topological charge (using numerical methods) with the "true"
value of its charge. We assume the same errors in energy and charge since
similar finite-difference methods are employed in calculating each of these
values. We have noticed that higher differences in the lattice spacing in each
space direction yield higher errors.   

The initial condition that we start with for minimization is the approximate
skyrme crystal of eight half-skyrmions, namely (\ref{eq:appsolcrystal1}) and
(\ref{eq:appsolcrystal2}). After being minimized, the period in each direction
for this initial configuration is then changed in a certain way (described in
the relevant section) and then re-minimized.

\subsubsection{From a 4-skyrmion to a square 2-wall}
\begin{figure} 
\centering
\includegraphics[width=0.8\textwidth]{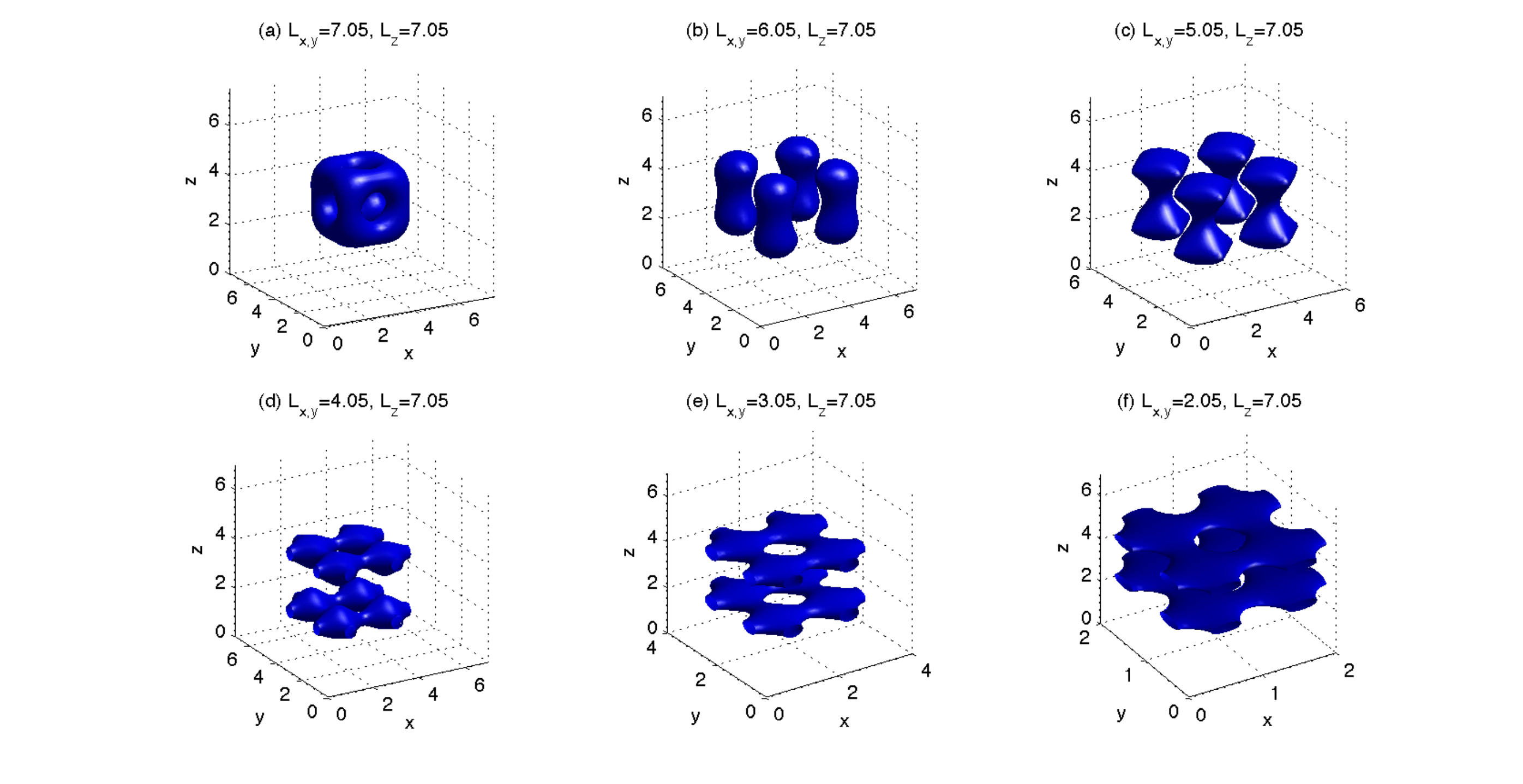} 
\caption{Energy density isosurfaces with surface value given by
$0.5*\mathcal{E}_{\mathrm{max}}$, where $\mathcal{E}_{\mathrm{max}}$ is the
maximum value of the energy density (all isosurfaces in subsequent figures have
the same surface value). The first isosurface corresponds to the $Q=4$ skyrmion
for the periods $L_{x,y,z}=7.05$. Each successive picture has $L_{x,y}$ reduced
by 1 and $L_{z}$ is kept constant at 7.05.} \label{fig:lxyreduce}
\end{figure}

The first case is the one for which the period in all three directions is large
for the initial configuration and then the $L_{x,y}$ periods are reduced
gradually. The initial period is given by $L_{x,y,z}=7.05$ (see Fig.
\ref{fig:lxyreduce}) and then reduce $L_{x}=L_{y}$ by 1 each time. 

We start by increasing the periods $L_{x}=L_{y}=L_{z}$ for the minimized Skyrme
crystal from $L_{x,y,z}=4.7$ to $7.05$ by increasing the value of $n_{x,y,z}$,
for a certain value of $h_{x,y,z}$, and thus producing Fig.1(a). Afterwards,
$n_{x,y,z}$ is kept constant and $L_{x,y,z}$ is reduced by $1$ each time.

One can see that the translation symmetries, $\mathcal{T}_{i+/-}$, (where
$i=\{1,2,3\}$ and in either direction $+/-$) of the high-density Skyrme crystal
are broken in Fig.1(a), which is the $Q=4$ skyrmion, whereas all reflection
symmetries, $\mathcal{R}_{i}$, three-fold rotations $\mathcal{R}^{3}_{1/2,+/-}$,
and four-fold rotations $\mathcal{R}^{4}_{i,+/-}$ remain unbroken --
characteristic of cubic symmetry. In Fig. 1(b)-(f), $\mathcal{T}_{1+/-}$ and
$\mathcal{T}_{2+/-}$ symmetries are regained (possibly due to the fact that we
are "squeezing" the configuration in these directions), whereas
$\mathcal{T}_{3+/-}$ remains broken only for Fig. 1(b) and Fig. 1(c), due to the
fact that these figures are not extended throughout the whole period in the
$z-$direction. The reflection symmetries $\mathcal{R}_{i}$ are unbroken and the
three-fold rotations $\mathcal{R}^{3}_{1/2,+/-}$ are broken in Fig. 1(b)-(f).
The four-fold rotations $\mathcal{R}^{4}_{3,+/-}$ remain unbroken throughout,
whereas $\mathcal{R}^{4}_{1/2,+/-}$ are broken in Fig. 1(b)-(f).

Note that a double Skyrme sheet configuration \cite{jslrsw} emerges in Fig.
1(d). The separation between the sheets, which is calculated by measuring the
distance between the energy density peaks as a function of $z$, remains constant
as $L_{x,y}$ is reduced further and $L_{z}$ is kept constant at $L_{z}=7.05$.
However, the energy decreases from a value of $E=4.36\pm0.09$ at
$L_{x,y,z}=7.05$ down to a minimum of $E=4.20\pm0.05$ at $L_{x,y}=4.05$ and,
finally, increases to $E=5.25\pm0.10$ at $L_{x,y}=2.05$.  

The preferred configuration for this Skyrme sheet is to have $L_{x}=L_{y}$. For
instance, if we fix $L_{y}=4.05$ and vary $L_{x}$ away from this value in either
direction, we notice that the energy increases.

\subsubsection{From square 2-walls to vortices}
We now start with a large $L_{z}$ period, which will be reduced, and keep
$L_{x,y}=4$ constant. The starting point is $L_{z}=7$, which is then reduced by
$1$ each time. In this section, $n_{x}=n_{y}=n_{z}$ are kept constant, producing
different $h_{z}$ values each time $L_{z}$ is changed. 

The initial condition (eqs. (\ref{eq:appsolcrystal1}) and
(\ref{eq:appsolcrystal2})) is first minimized for the periods $L_{x,y}=4$ and
$L_{z}=7$, producing the double Skyrme sheet configuration seen in Fig. 2(a).
The only broken symmetries associated with this configuration, which is seen to
persist in Fig. 2(b) and Fig. 2(c), are  $\mathcal{R}^{4}_{1/2,+/-}$ and
$\mathcal{R}^{3}_{1/2,+/-}$. These are regained in Fig. 2(d), which is a
(non-minimal) Skyrme crystal configuration. As the $L_{z}$ period is decreased
further, we notice the appearance of vortex-like structures, which are periodic
in the $z-$direction (Fig. 2(e) and Fig. 2(f)). The symmetries associated with
these configurations are the same as those in Figs. 2(a)-(c) and, in fact, the
translation symmetry in the $z-$direction, $\mathcal{T}_{3+/-}$, becomes
continuous for Fig. 2(f), since there is no longer any noticeable
$z-$dependence. This ties into the subject of Skyrme chains, which has been
explored in \cite{Harland:2008eu}.

The energy of the isosurfaces decreases from a value of $E=4.22\pm0.05$ at
$L_{z}=7$, down to a minimum of $E=4.17\pm0.04$ at $L_{z}=5$ (where the
isosurfaces are still in the form of a double square wall), then back up to
$E=4.89\pm0.05$ at $L_{z}=2$.

\begin{figure} 
\centering
\includegraphics[width=0.8\textwidth]{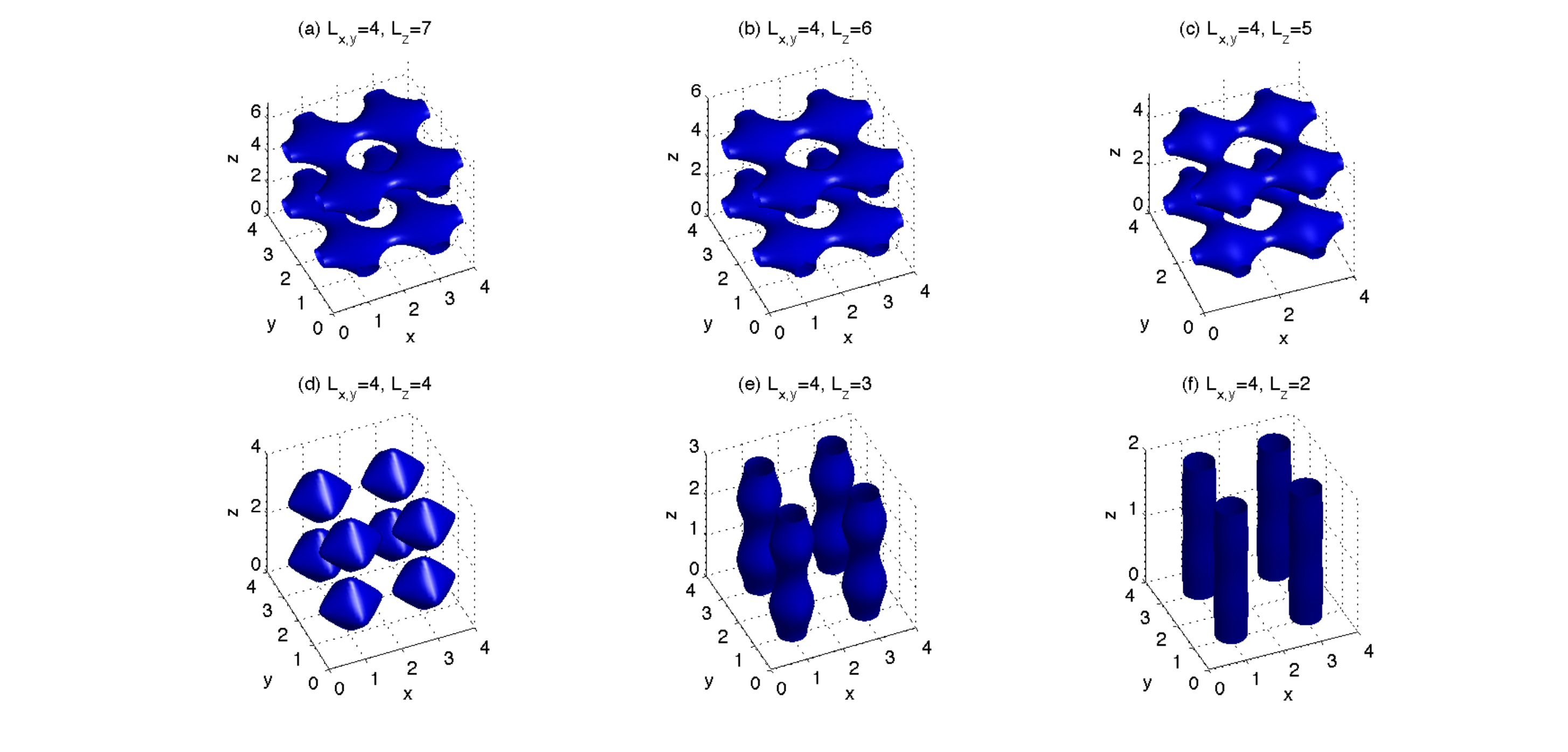} 
\caption{Energy density isosurfaces corresponding to the $Q=4$ double Skyrme
sheet configuration, where $L_{x,y}=4$ remains fixed and where $L_{z}=7$ is
reduced by 1 in each successive picture, transforming the double skyrme sheet
into a (non-minimal) Skyrme crystal and thereafter into a parallel 4-vortex
structure.} \label{fig:lxyconstantlzreduce}
\end{figure}

\subsubsection{From square 2-walls to vortices, via crystal}
\begin{figure} 
\centering
\includegraphics[trim=0cm 1.0cm 0cm 0cm, clip=true,
width=0.8\textwidth]{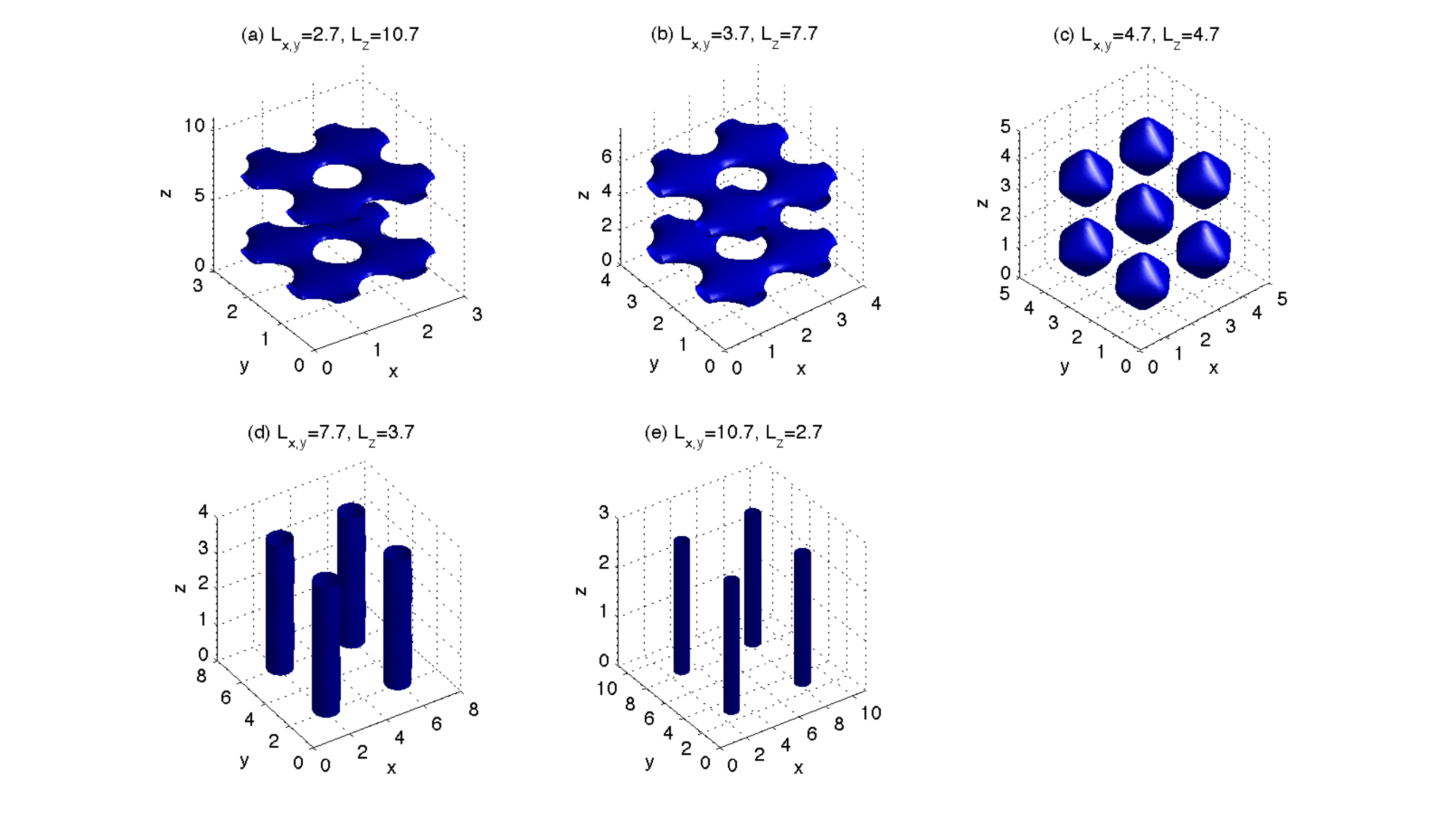} 
\caption{Energy density isosurfaces, which show a transition from a pair of
Skyrme sheets to a 4-vortex configuration. Subsequent pictures have $L_{x,y}$
and $L_{z}$ changing in such a way that their values are swapped halfway through
the transition.} \label{fig:lxyexpandlzreduce}
\end{figure}

In this section, we show how one can transform a double Skyrme sheet
configuration into the 4-vortex configuration discussed in the previous section,
by changing both $L_{x}=L_{y}$ and $L_{z}$ (rather than just $L_{z}$), and going
through an intermediate, minimal, Skyrme crystal state, as can be seen in Fig.
\ref{fig:lxyexpandlzreduce}. The initial period is $L_{x,y}=2.7$ and
$L_{z}=10.7$. The former is then increased by 1 and the latter decreased by 3,
two consecutive times, producing Figs. 3(b)-(c). The changes in the periods are
then swapped, increasing $L_{x,y}$ by 3 and decreasing $L_{z}$ by 1, for two
consecutive times, producing Figs. 3(d)-(e). Here, $n_{x}=n_{y}=n_{z}$ are kept
constant, producing different $h_{x}=h_{y}$ and $h_{z}$ values, each time the
periods are changed.

We start by minimizing the approximate Skyrme crystal (eqs.
(\ref{eq:appsolcrystal1}) and (\ref{eq:appsolcrystal2})) for the periods
$L_{x,y}=2.7$ and $L_{z}=10.7$, producing the double Skyrme sheets discussed in
the previous sections (with the same symmetries) in Fig. 3(a). As the periods
are changed, as described above, the Skyrme sheets still persist in Fig. 3(b),
and change into the (minimal-energy) Skyrme crystal in Fig. 3(c) (with all its
associated symmetries as described in the introduction). As $L_{x,y}$ increase
and $L_{z}$ decreases further, the Skyrme crystal changes into a 4-vortex
structure, Figs. 3(d)-(e), which have less of a $z-$dependence than the ones in
Figs. 2(e)-(f), respectively.  

The Skyrme crystal has the minimum energy, with $E=4.13\pm0.04$, followed by the
square 2-walls at $L_{x,y}=3.7$ and $L_{z}=7.7$ with $E=4.26\pm0.06$. The final
picture, the 4-vortex configuration with the smallest $L_{z}-$value, has the
highest energy, with $E=4.77\pm0.17$, which follows from the fact that Skyrme
chains ``prefer'' to be closer together in the $x,y-$directions
\cite{Harland:2008eu}.

Thus far, we have changed the periods in the three space directions in different
ways - by decreasing $L_{x,y}$ (or keeping it constant) and keeping $L_{z}$
constant (or decreasing it) or changing all three periods at different rates at
the same time. We now turn to the case where we increase (or decrease) all three
periods simultaneously at the same rate.  

\subsubsection{From the Skyrme crystal to the $Q=4$ skyrmion}
\label{sec:alphagt0}
It turns out that an interesting feature of the $Q=4$ system is uncovered when
one starts from the minimal-energy Skyrme crystal configuration at
$L_{x,y,z}=4.7$ and then increases the periods simultaneously until the
individual half-skyrmions coalesce.

\begin{figure}[!h]
\centering
\includegraphics[trim=2cm 4cm 0cm 3cm, clip=true,
totalheight=0.19\textheight]{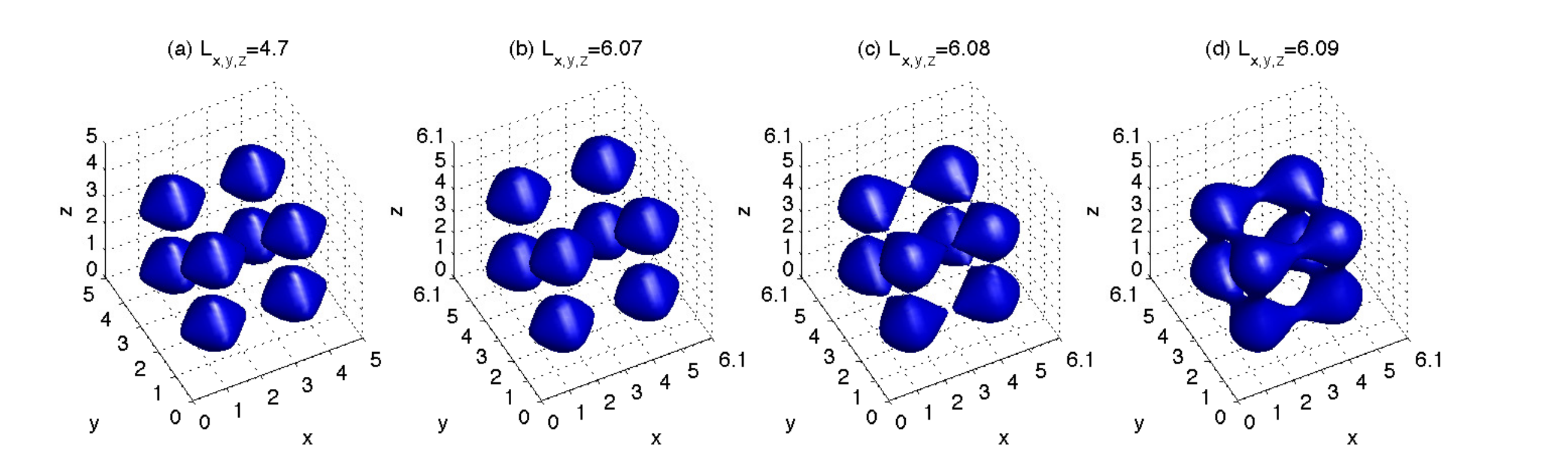} 
\caption{The energy density isosurface of the $Q=4$ Skyrme crystal at
$L_{x,y,z}=4.7$. The periods are then increased gradually from $L_{x,y,z}=6.07$
to $L_{x,y,z}=6.09$.} \label{fig:lxyzexpand}
\end{figure}

\begin{figure}[!h]
\centering
\includegraphics[trim=2cm 0.7cm 0cm 0cm, clip=true,
totalheight=0.18\textheight]{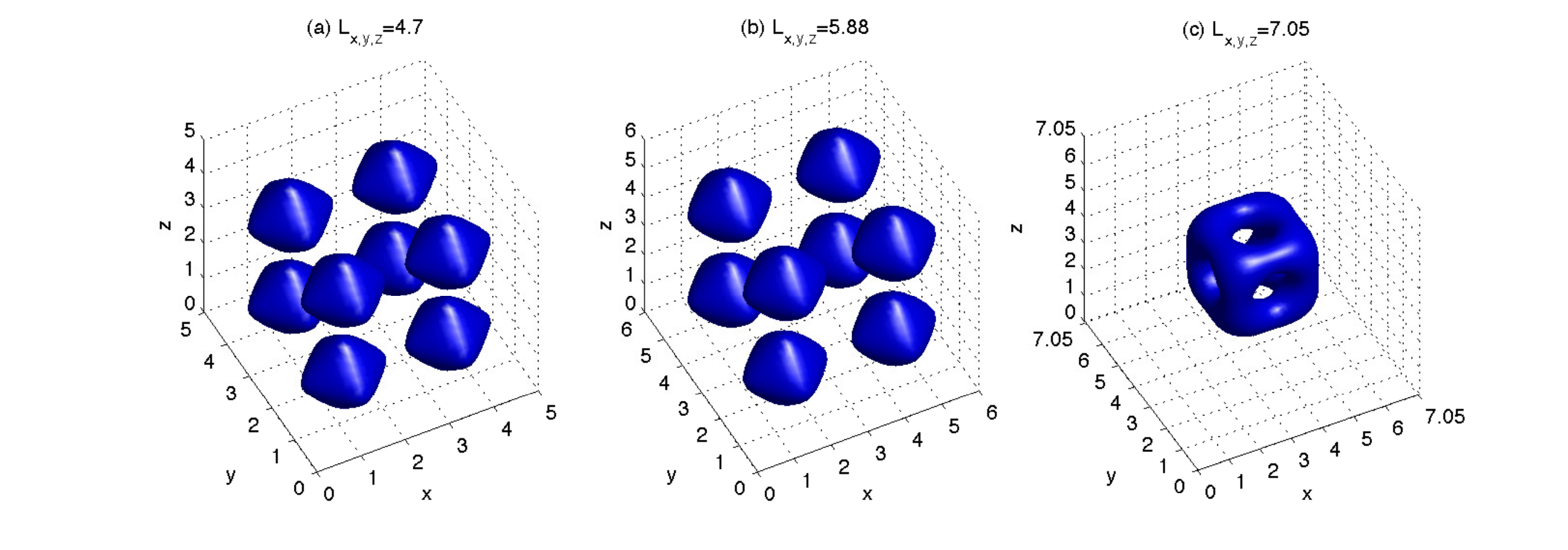} 
\caption{The first figure corresponds to the energy density isosurface of the
Skyrme crystal for the periods $L_{x,y,z}=4.7$, the second figure has periods
$L_{x,y,z}=5.88$, and the third one has $L_{x,y,z}=7.05$ } \label{fig:cube}
\end{figure}

These half-skyrmions clump in a sudden fashion when one perturbs the system in a
certain way. We perturbed it by removing one lattice site through the middle of
the configuration in all three directions, thereby ``squeezing'' the
half-skyrmions together in all directions. Therefore, in order to see them
coalescing, one needs to increase the periods very gradually. This appears to
happen in the range $L_{x,y,z}=6.07-6.09$ as can be seen in Fig.
\ref{fig:lxyzexpand}, where the first picture has periods $L_{x,y,z}=4.7$ and
the following three correspond to the range just mentioned.\footnote{It is worth
mentioning that the value of this range changes when the bin size is changed -
we used the value $n_{x,y,z}=32$. However, the general feature that the
half-skyrmions coalesce rapidly remains the same.} Compare this with Fig.
\ref{fig:cube}, where the periods are increased in larger steps.

\begin{figure}[!h]
\centering
\includegraphics[trim=0cm 0.5cm 0cm 1cm, clip=true,width=0.73\textwidth]{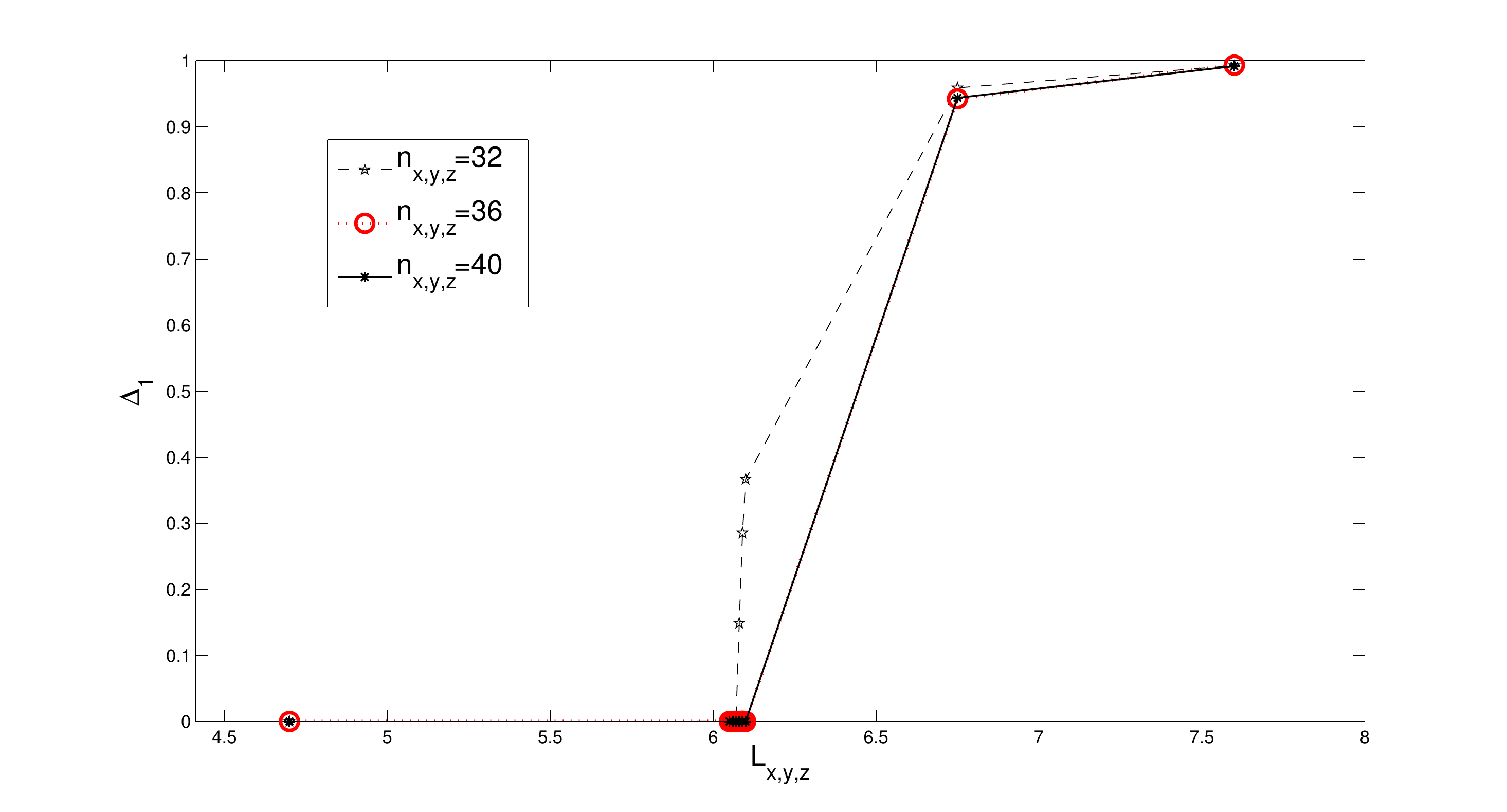} 
\caption{Difference in the value of the energy density of the Skyrme fields
under an $L_{x}/2-$translation in the $x-$direction divided by
$\mathcal{E}_{\mathrm{max}}$, for different values of the period $L_{x,y,z}$.
Note that as the bin size decreases (or as $n$ increases), the plot begins to
look more like a step function.} \label{fig:asymmetries}
\end{figure}

The sudden merging of the half-skyrmions can be visualized in a different
fashion by taking the maximum value of the difference in the energy densities of
the fields $\mathcal{E}[\Phi_{\beta}(x^{j})]$ under a certain symmetry
transformation, in this case a translation of $L_{x}/2$ in the $x-$direction,
which is half the size of the fundamental cell, and dividing by the maximum
value of the energy density, i.e.
$\Delta_{1}=(\mathcal{E}[\Phi_{\beta}(x^{j})]-\mathcal{E}[\Phi_{\beta}(x^{j'})]
)_{\mathrm{max}}/\mathcal{E}_{\mathrm{max}}$, where $x^{j'}=x^{j}-L_{x}/2$. We
then plot this as a function of the period $L_{x,y,z}$ - the reason we do this
is that, as soon as the half-skyrmions start to coalesce, they will no longer be
$L_{x}/2-$periodic. For the Skyrme crystal, this difference is seen to be
essentially zero and it then starts to increase as the half-skyrmions begin to
coalesce as can be seen in Fig. \ref{fig:asymmetries}

This jump in the asymmetries of the crystal is analogous to a phase transition
in thermodynamics. There is a sudden transition from a crystalline phase, which
has more symmetries, such as chiral $\mathop{\rm SO}(4)$ symmetries for the Skyrme crystal, to a phase with less symmetries, such as $\mathop{\rm SO}(3)$ isospin symmetries for the $Q=4$ cubic-shaped skyrmion. We expect the transition to become more pronounced as the number of bins increases -- as can be seen in Fig. 6, when $n_{x,y,z}$ is increased from 32 to 36, which starts to resemble a step function. It should be noted that the $\Delta_{1}$ values are independent of which $L_{x,y,z}$ value one starts with, which hints to a lack of hysteresis in the system. Extending the analogy with phase transitions, this lack of hysteresis strongly suggests a second-order phase transition (\emph{cf.} \cite{visintin}), with no latent heat, and with the period $L_{x,y,z}$ as the order parameter.
  
%%%%%%%%%%%%%%%%%%%%%%%%%%%%%%%%%%%%%%%%%%%%%%%%%%%%%%
\section{Concluding Remarks}

We have seen that changing the periods of the Skyrme crystal, away from the
values which minimize its energy, has allowed us to find new configurations --
such as the double square lattice of "Skyrme sheets" -- as well as unearthing
unforeseen connections with Skyrme chains. The Skyrme sheets are produced when
the period $L_{x,y}$ is small compared with $L_{z}$. Our results \cite{jslrsw}
have shown that these are more energetically favourable than the double
hexagonal case, which makes them a prime candidate for use as building blocks
for large-charge configurations. If one then goes to the other limit of large
$L_{x,y}$ and small $L_{z}$ one gets a series of four vortex-like charge $Q=1$
objects, which is related to the subject of Skyrme chains \cite{Harland:2008eu}.
As the periods are changed and we go through these different configurations, we
have seen that these generally merge and separate in the directions in which one
is changing the period, for example, the merging of the half-skyrmions in the
$z$ direction as the period is decreased along this direction.

We have also noticed that the phase transition from the Skyrme crystal to a charge $Q=4$ skyrmion is likely to be second-order, due to the lack of hysteresis in the system, and with an order parameter given by the period $L_{x,y,z}$. 

Note that the deformations considered here have all been along one or more of the edges of the Skyrme crystal, keeping it rectangular. An interesting extension, which we plan to investigate, would be to see if any new structures emerge when one applies more general affine transformations, such as a stretch along a diagonal of the crystal. 

\bigskip\noindent{\bf Acknowledgment.}
I would like to thank Professor Richard Ward for helping me with the editing process, Professor Nick Manton for the useful and interesting discussions about hysteresis, and Carlos Silva Platt for his support. 
%%%%%%%%%%%%%%%%%%%%%%%%%%%%%%%%%%%%%%%%%%%%%%%%%%%%%%

\end{document}